\documentclass[preprint,journal]{vgtc}       

\usepackage{mathptmx}
\usepackage{graphicx}
\usepackage{times}
\usepackage[dvipsnames]{color}

\definecolor{mygray}{gray}{0.6}

\onlineid{0}

\vgtccategory{Research}

\vgtcinsertpkg

\preprinttext{arxiv version}

\title{Anchored in a Data Storm: How Anchoring Bias Can Affect User Strategy, Confidence, and Decisions in Visual Analytics}

\author{Ryan Wesslen, Sashank Santhanam, Alireza Karduni, Isaac Cho, Samira Shaikh, and Wenwen Dou}

\author{Ryan Wesslen\thanks{e-mail:rwesslen@uncc.edu} %
\and Sashank Santhanam
\and Alireza Karduni
\and Isaac Cho %
\and Samira Shaikh
\and Wenwen Dou\thanks{e-mail:wdou1@uncc.edu}}
\affiliation{\scriptsize University of North Carolina at Charlotte}

\shortauthortitle{Biv \MakeLowercase{\textit{et al.}}: Anchored in Data Storm}

\abstract{Cognitive biases have been shown to lead to faulty decision-making. Recent research has demonstrated that the effect of cognitive biases, anchoring bias in particular, transfers to information visualization and visual analytics. However, it is still unclear how users of visual interfaces can be anchored and the impact of anchoring on user performance and decision-making process. To investigate, we performed two rounds of between-subjects, in-laboratory experiments with 94 participants to analyze the effect of visual anchors and strategy cues in decision-making with a visual analytic system that employs coordinated multiple view design. The decision-making task is identifying misinformation from Twitter news accounts. Participants were randomly assigned one of three treatment groups (including control) in which participant training processes were modified. Our findings reveal that strategy cues and visual anchors (scenario videos) can significantly affect user activity, speed, confidence, and, under certain circumstances, accuracy. We discuss the implications of our experiment results on training users how to use a newly developed visual interface. We call for more careful consideration into how visualization designers and researchers train users to avoid unintentionally anchoring users and thus affecting the end result.
} 


\CCScatlist{ 
 \CCScat{K.6.1}{Management of Computing and Information Systems}%
{Project and People Management}{Life Cycle};
 \CCScat{K.7.m}{The Computing Profession}{Miscellaneous}{Ethics}
}

\begin{document}


\firstsection{Introduction}

\maketitle

An emerging topic within the Visual Analytics (VA) community focuses on understanding the impact of cognitive biases on the analysis process aided by visual analytic systems. VA combines automated analysis techniques with interactive visualizations to facilitate human decision-making processes on large and complex data. One of the many factors that contribute to an effective VA system is the support of \emph{exploratory visual analysis} \cite{munzner2014visualization, keim2002information}. Many VA systems designed to support exploration often employ coordinated multiple views (CMV) to present various aspects of the underlying data and analysis results. These VA systems offer the flexibility of devising different strategies to solve problems the systems are designed to address. The strategies can materialize in how the users interact with VA systems, including relying on all or a subset of the coordinated views, as well as the perceived importance of different views. As a result, during exploratory visual analysis, users are faced with a potentially overwhelming array of choices while being constrained by limited cognitive resources and uncertainty. For instance, users need to decide on the view to start their analysis, where to go next within the visual interface, how to interpret and synthesize patterns seen from multiple views, as well as the combination of views to rely on for their decisions. Such exploratory visual analysis processes are prone to cognitive biases \cite{sacha2016role}. 

Cognitive biases are rules of thumbs or heuristics that aid in decision-making tasks and allow users to reach decisions with relative speed \cite{tversky1974judgment}. Cognitive biases have been shown to affect decision-making processes in predictably faulty ways that can result in sub-optimal solutions when information is discounted, misinterpreted, or ignored \cite{tversky1974judgment}. One cognitive bias   particularly relevant to exploratory visual analysis with VA systems is anchoring bias. It refers to the human tendency to rely too heavily on one and most likely the first piece of information offered (the “anchor”) when making decisions \cite{kahneman201636}. Numerous studies from the fields of psychology and behavioral economics have analyzed the effect of numerical anchors, showing difficulties for participants to adjust away from the initial numerical value (anchor) provided \cite{kahneman201636}. In prior work, we demonstrated that the anchoring effect transfers to VA; specifically we described ``visual anchoring'', which refers to the over reliance on a single or subset of views during exploratory visual analysis with VA systems that employ CMV design \cite{cho2017anchoring}. As one of the first studies on the effect of anchoring bias in VA, our prior work is situated in an open-ended task (identifying protest-related events from social media data). Therefore, we analyzed the impact of visual anchors on the analysis paths but not on user performance (due to the absence of ground truth events). Moreover, no comparisons were made between visually anchored groups against a control group (no visual anchor given). 

The experiments presented in this paper are designed to analyze the impact of visual anchors on a variety of quantitative metrics including accuracy, time, user interactions, data coverage, as well as ways that users can be visually anchored. To situate our study in a real-world reasoning task, we chose the application of misinformation investigation of social media news accounts. Recently, the topic of combating misinformation has received much attention in many fields including data mining, journalism, and computational social science \cite{lazer2018science, pennycook2017prior, shu2017fake}. While a variety of computational techniques have been explored, some scholars have also called for the need to study misinformation in randomized, controlled laboratory experiments \cite{vosoughi2018spread}. In our study, we use a visual analytics tool designed for investigating misinformation. We design multiple treatments/conditions in order to analyze the effect of visual anchors while participants performing the task of evaluating the veracity of news accounts on Twitter. 


Our work makes the following salient contributions:
\begin{enumerate}
 \setlength\itemsep{0em}
    \item The design of experiments situated in a task of making decisions about the veracity of news media accounts on Twitter using a visual interface designed for investigating misinformation, as two rounds of between-subjects in-laboratory experiments ($n = 94$) to test the effect of visual anchoring in decision-making.

    \item Careful integration of psychology literature on ways a user can be anchored in exploratory visual analysis to reveal the effect of anchoring with strategies or cues. 
    \item Quantitative analysis performed on a range of factors that effect anchoring bias in VA to reveal findings on how visual anchors impact user performance and data coverage as well as user confidence on their decisions. 

%

\end{enumerate}
Understanding the effect of various cognitive biases in visual analysis and how the biases are reflected in the analysis process with a VA system serve as an important first step to raising awareness and ultimately mitigating cognitive biases in visual analysis.
At the end of the paper, we connect findings from our experiments to practices of interacting with participants on a newly designed visual analytic system. 
The findings of our user study shed more light on how and when anchoring bias could occur when using visual analytic interfaces and call for more careful consideration when introducing a visual interface to end-users or designing tutorials of a visual analytic system.  
\section{Related Work}

We summarize the current research effort on cognitive biases in visualization into two categories: holistic approaches aiming at framework and metrics for studying cognitive biases in visualization research, and empirical studies of how a certain type of cognitive bias manifests and impacts the analysis process facilitated by visual analytic systems. 
We also review literature that motivated our experiment design and research questions. 

\subsection{Cognitive Bias and Anchoring in VA}

Cognitive Bias are systematic patterns in judgments that deviate from rationality due to a variety of factors (e.g., unfamiliarity, too much information, quick decision-making). Psychologists and social scientists have followed the seminal work of Tversky and Kahneman \cite{tversky1974judgment} to investigate a variety of cognitive biases in a variety of applications \cite{blackwell2015positive}. Recently, visual analytics community has started to explore the role cognitive biases play in decision-making processes \cite{valdez2017framework, wall2017warning, valdez2018priming}. In this paper, we follow our previous empirical study \cite{cho2017anchoring} to further investigate the role of anchoring bias within a CMV system.

Our precedent for providing visual anchors and strategy cues in the experiment design is rooted in the fundamental literature from psychology \cite{rajsic2015confirmation,amer2017biasing,wright2017argument,lieder2017anchoring,shaffer2015manual,ellis2015decision,bonaretti2017cognitive}. To illustrate, research has demonstrated that users preferred to devote attention to stimuli that matched a given hypothesis or template, even in the presence of alternate, more optimal strategies \cite{rajsic2015confirmation}. The work of \cite{amer2017biasing} designed experiments in which participants were given explicit and implicit spatio-temporal cues in a visual event coding task and found systematic effects of the explicit and implicit cues on users' attention within the visual analytic system and how these cues affected processing of information. 

\subsection{Reviews on Practices of Evaluating Visualization}
The findings from our experiments are relevant to a critical step, providing training and tutorial, during visualization evaluation with human subject. Therefore, we briefly summarize existing work on the theories and practices of visualization evaluation and highlight which step in evaluating visualization that our findings can inform. Visualization evaluation has always been an integral part of research in the VIS community \cite{Lam:2012, VisSurvey:2013, Plaisant:2004}. By surveying 850 papers from the InfoVis and VAST venues, Lam et al. identified seven evaluation scenarios in order to guide practitioners to design effective user studies \cite{Lam:2012}. Two out of the seven scenarios specific to understanding visualizations involve directly interacting with participants. The two scenarios, namely evaluating user performance and evaluating user experience, are among the most frequently used evaluation techniques. Although the scenarios provide great guidance on the selection of appropriate questions and goals for user evaluation, there are several challenges when carrying out the evaluation. Such challenges include short duration of many study periods \cite{Plaisant:2004}, insufficient number of study measures \cite{Plaisant:2004}, and possibly inadequate training of participants \cite{BELIV:2008}. 

Another recent survey of visualization evaluation practices from the Vis Community echoed these challenges and highlighted that many publications need to observe more evaluation reporting rigor by providing important methodological details \cite{VisSurvey:2013}. Based on the findings from our experiments, we argue that reporting how the participants were trained (by experimenters, with or without a script, training videos, example strategies to complete the task, etc.) should be consistently reported.
Our results show that the training can have a significant impact on how participants use of the interface, as well as their performance and their perceived confidence on completing the task(s).

\begin{figure}[t]
  \centering
    \includegraphics[width=1.0\columnwidth]{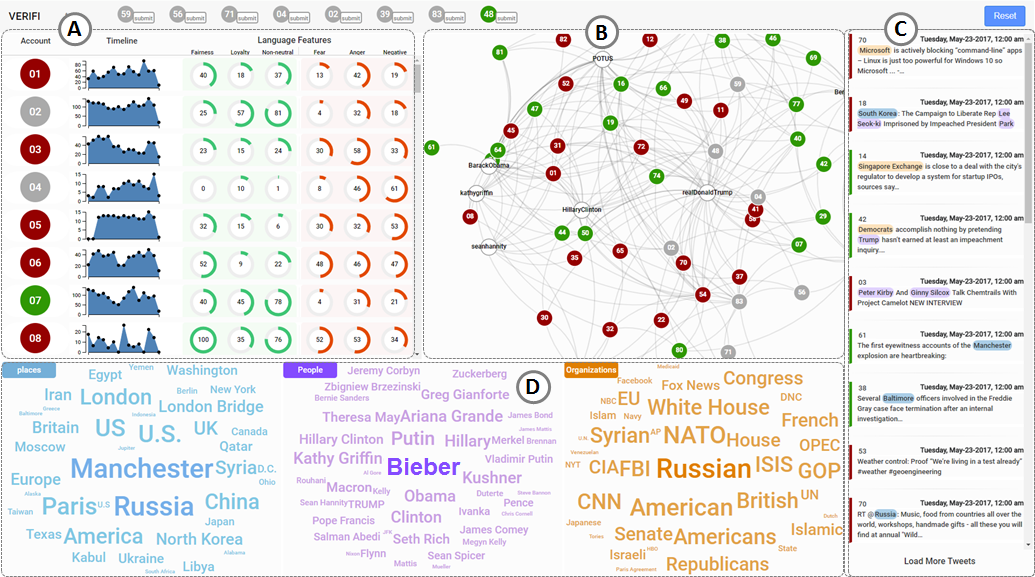}
  \caption{Screenshot of Verifi. Verifi is comprised of four views: (A) Accounts View, (B) Social Network View, (C) Tweets Panel View, and (D) Entities View. Progress Bar and Form Submit buttons are at the top.}
  \label{fig:verifi}
\end{figure}

\begin{figure}[ht]
  \centering
    \includegraphics[width=1.0\columnwidth]{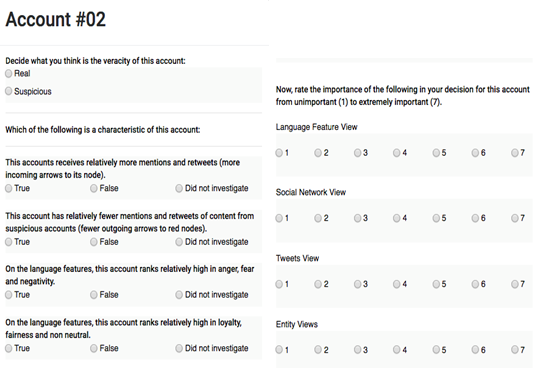}
  \caption{Form Submit view of Verifi for Account \#02. This pop-up provides an interface for the user decisions and feedback per account (e.g., strategy cues use, view importance, and open-ended feedback (not shown).)}
  \label{fig:form-submit}
\end{figure}

\section{Method}

In this section, we first review our visual analytics system, Verifi, and then outline of research questions.

\subsection{The Verifi System}

For our study, we use Verifi \cite{karduni2018icwsm} (Fig. \ref{fig:verifi}), an interactive, coordinated-multiple views system for identifying Twitter news accounts suspected of spreading misinformation. Verifi includes four main views: Social Network, Accounts, Tweet Panel, and Entities. Each view provides users with different factors that have shown to be important in detecting misinformation \cite{volkova2017separating}. The Social Network and the Accounts views are the two primary views that serve as the two visual anchors. The Entity View and Tweet Panel are secondary views.

The data includes 82 Twitter news accounts anonymized by name but annotated with color labels indicating whether they are source of misinformation (red), real news outlets (green), or require the users' decision (grey). The annotations are based on multiple third-party sources \footnote{Suspicious accounts are based on four websites as provided in \cite{volkova2017separating}. 31 real news accounts are provided through the following links: https://tinyurl.com/yctvve9h and https://tinyurl.com/k3z9w2b}. Each user's \textbf{task} is to make a decision on the veracity (real or suspected of spreading misinformation) for eight grey accounts within a one-hour session. Building on our prior study, these eight grey accounts have been qualitatively selected to provide a range of difficulties as well as consistent and inconsistent information to challenge users in their decision-making processes \cite{karduni2018icwsm}. Table \ref{fig:account} provides the anonymized names of the eight gray accounts (four real and four suspicious according to third party sources) along with a brief description. 

\subsection{Overhauled Experiment Design}
In our past study \cite{cho2017anchoring}, we investigated anchoring bias within a visual analytic system (CrystalBall \cite{cho2017crystalball}) that employs CMV design for event detection in Twitter. In this paper, we have revamped the experiment design in three ways. 
(1) We collect direct input on users' decision-making process, we included a form submission view (Figure \ref{fig:form-submit}) to explicitly capture the precise moment when users make a decision about misinformation, and allow users to directly rate the helpfulness of the strategy cues to each decision within the system. 
(2) We provide explicit strategy cues (in the form of written cues and reinforced in the training videos) as a second treatment condition for each primary view in Verifi \cite{karduni2018icwsm}. In this way, we can control for the role as well as measure users' evaluation of that strategy for each decision. 
(3) We quantitatively evaluate the impact of visual anchors and strategy cues on users' performance, we designed the experimental task in a way that the users' answers can be measured against ground truth. The task in our past study with CrystalBall was exploratory in nature, thus we couldn't measure users' accuracy.

\subsection{Research Questions}
We seek to investigate how users may be anchored on different \textit{views} in a CMV system and how they might be anchored on specific interaction \textit{strategies} based upon the training given to them. Further, how does anchoring affect user performance, confidence and data coverage? Accordingly, our two main research questions (RQs) are:

\textbf{RQ1:} What is the effect of visual anchors and strategy cues on participant performance (i.e., correctness, speed, and confidence) and ratings (e.g., view importance and strategy usage)?

\textbf{RQ2:} Can users' analysis process (e.g., interaction logs) be linked to participant performance outcomes to infer user strategies?

To analyze RQ1, we use univariate statistical tests (e.g., one-way ANOVA, Kruskal-Wallis Rank Sum) as well as multivariate regression (e.g., linear, logistic) to consider the effect among additional independent variables (e.g., account or time).  For RQ2, we use feature extraction to obtain time spent in each view and coverage metrics \cite{wall2017warning} to understand user strategies through their interaction logs. After isolating features that measure primary actions, we cluster users based on their usage patterns and validate against their responses to identify unique behaviors attributed to each group. Using visual analytics, we explore user-level interactions by these clusters to identify salient behaviors and strategies. 

\begin{table}[t]
  \centering
    \includegraphics[width=0.9\columnwidth]{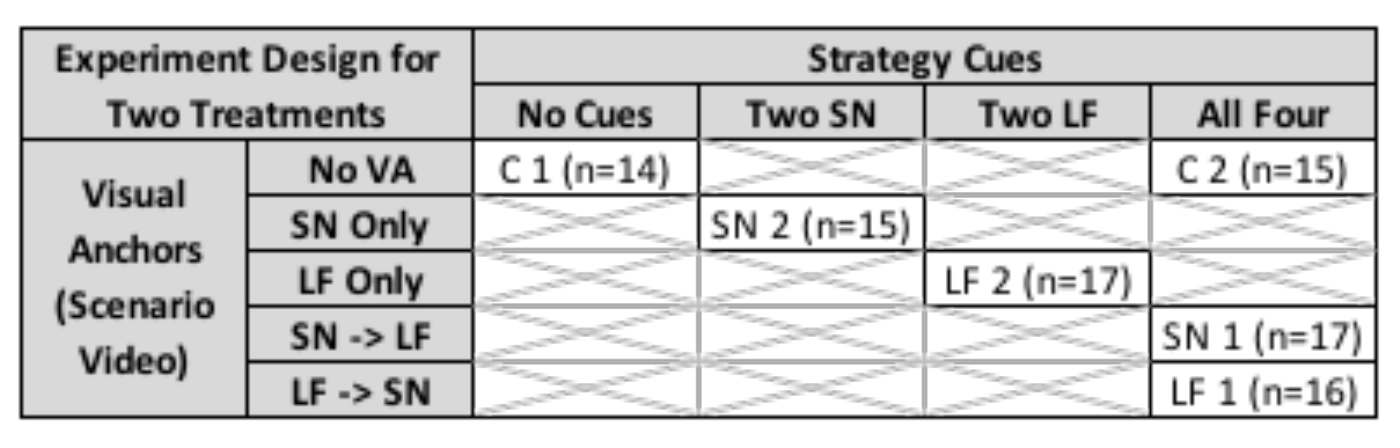}
  \caption{Experiment design for two treatments (columns) across two study rounds. Each cell represents a study-treatment group per one of two treatments: strategy cues (columns) or visual anchors (rows). C = Control group, SN = Social Network, LF = Language Features.}
  \label{tab:treatments}
\end{table}

\section{Experiment}
\label{section:experiment}

To analyze the effects of visual anchors and strategy cues in decision-making, we performed two rounds of between-subjects, in-laboratory experiments. Each user's task is to make a decision on the veracity (real or suspicious) of the eight grey accounts (see Figure \ref{fig:account}). Users could make their decisions at any time by entering the Form Submit view (Figure \ref{fig:form-submit}) in the Progress Bar view for each account. To control for learning effects, we randomized the order the accounts were presented in the Progress Bar per unique user ID. 

Following established psychology experiment design, we explicitly devised \textit{strategy cues} to present to users as part of our experiment condition. Each strategy cue aligns to one of two primary views in the Verifi VA system: Accounts view (L) and Social Network view (S). The Accounts view presents information about how each Twitter news account score on the language features, such as fairness, loyalty, anger, and fear. While the Social Network shows account connections through retweets and mentions. The cues were given as a piece of paper to users. The text of the cues are:

\textbf{Cue 1L}: ``On the language measures, real news accounts tend to show a higher ranking in loyalty, fairness, and non-neutral.''

\textbf{Cue 2L}: ``On the language measures, real news accounts tend to show a lower ranking in anger, fear and negativity.''

\textbf{Cue 1S}: ``In the social network graph, real news accounts are less likely to mention and retweet content from suspicious accounts (fewer outgoing arrows to red nodes).''

\textbf{Cue 2S}: ``In the social network graph, real news accounts tend to receive more mentions and retweets (more incoming arrows to their nodes).''

\begin{table}[t]
  \centering
    \includegraphics[width=0.9\columnwidth]{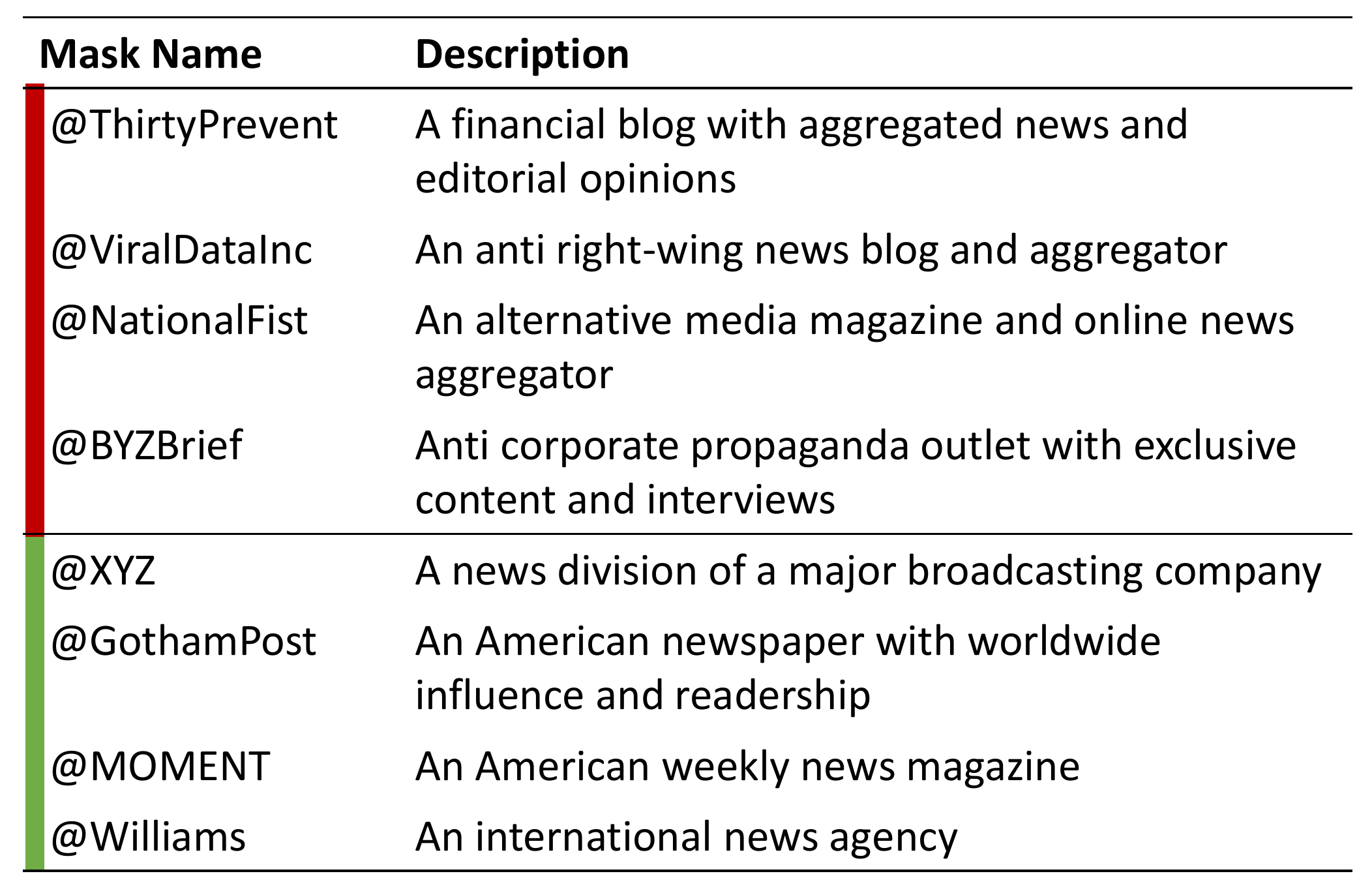}
  \caption{Eight Twitter news accounts selected for users' decisions (i.e., grey accounts in the interface). The names have been masked due to institutional concerns.} 
  \label{fig:account}
\end{table}

\begin{table*}[t]
  \centering
    \includegraphics[width=1.0\textwidth]{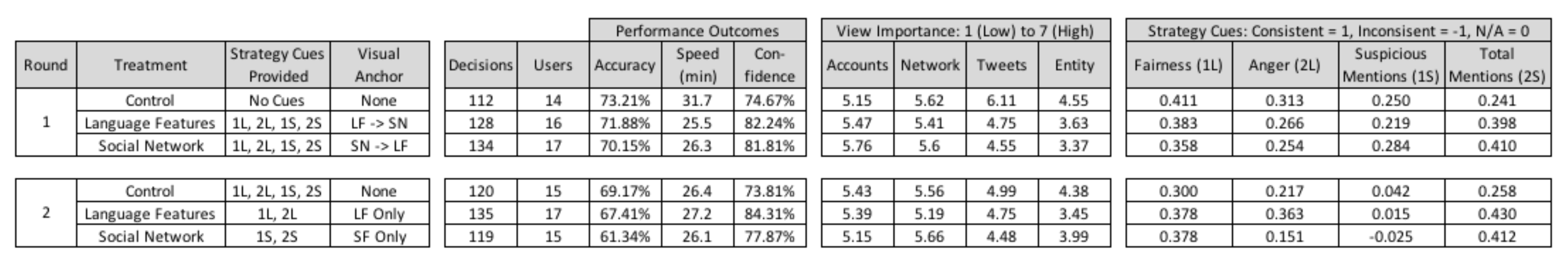}
  \caption{User response descriptive (mean) statistics by treatment group and round. Each row represents a different treatment group for the two rounds. The Strategy Cues and Visual Anchors provide different treatments per group. }
  \label{fig:summary}
\end{table*}
\vspace{0em}
\subsection{Descriptive Statistics}

In \textbf{Round 1}, we examine whether the starting point in the strategy provided to users would anchor them on a particular view, since certain psychological studies suggest that users are usually anchored on the \textit{first} piece of information \cite{tversky1974judgment}. Thus, Round 1 explores the role of dual strategies (social network or language features), reversing the order in which participants in a given treatment group are presented with the visual anchors and strategy cues. The control group participants in this round are given neither a visual anchor nor strategy cues. Round 1 took place in December, 2017. The findings from Round 1 justifies a follow-up study to tease out the effect of visual anchoring not only by the starting point in the training strategy, as well as the individual effect of visual anchors and strategy cues. In \textbf{Round 2}, we provided more focused treatments (i.e., only one set of view-based strategy cues and related visual anchor) along with a variant control group (i.e., all four cues, but no visual anchor). Round 2 took place in February, 2018. Combined, both studies enable a full investigation of two treatment mechanisms. Table \ref{tab:treatments} provides the treatment groups per round. 

The entire user session lasted around one hour and included pre- and post-questionnaires. The actual task with the visual interface was capped at 45 minutes and averaged slightly less than 30 minutes (M = 27.1 minutes, SD = 7.524, 25\% percentile = 20.98, 75\% percentile = 31.88). Each session is identified through user's participant ID and interactions like clicks, hovers, and scrolls were tracked and saved in our database. Computer specifications (browser, output/zoom) were controlled for to avoid them as confounding factors. 


Our study included 94 participants divided into two rounds, each with 47 participants.\footnote{In the first round, we excluded an additional 15 users (S1 - S15) but Verifi's mechanism to record responses failed during their user session. Hence, we could not record their decisions.} Users could participate in only one round, not both. The gender distribution was 68\% male and 32\% female. Users' ages were between 21 and 56 (M = 28.67). A majority of users were pursuing their Master's (88\%), followed by Undergraduate (5\%), Other (5\%), and Ph.D. (1\%). Students were recruited through extra credit incentives offered in one of six courses. The courses included Visual Analytics (n = 40), Natural Language Processing (n = 25), Advanced Business Analytics (n = 14),  Human Behavior Modeling (n = 6), Applied Machine Learning (n = 6), and Social Media Communications (n = 3). 

\subsection{Round 1: Dual Anchors and Cues}

\subsubsection{Experiment Setup}

In Round 1, we recruited 47 participants who were randomly assigned one of three treatment groups. As shown in Table \ref{tab:treatments}, the two treatment groups were provided all four strategy cues (introduced in Section \ref{section:experiment}) while the control group received no cues. The treatment groups differed by their visual anchor; the order of each view depended on the group. For example, the Social Network (SN) group's scenario video \textbf{starts} the investigation in the Social Network View and arrives at a conclusion of an example account being real or suspicious, this finding is then reinforced by investigation in the Language Features (LF) view. Similarly, the LF group's scenario video \textbf{starts} the investigation in the Language Features view. The two treatment group received the same information, only the order (LF or SN) was swapped. 

\subsubsection{Experiment Results - RQ1}

\textbf{Univariate analysis.} In Round 1, we find evidence that visual anchors and strategy cues had an effect on users' confidence and, weakly, overall time spent during analysis. Table \ref{tab:rq1} provides the respective univariate statistical tests for the treatment effects on each outcome per round. We find that confidence is significantly different among treatments in Round 1, driven by the much lower confidence in the Control group. Alternatively, we find that Time Spent is weakly significant, again driven by a much longer average session of the Control group (M = 31.7 minutes) than the two treatments (see Figure \ref{fig:summary}).

Table \ref{tab:view-importance} provides Kruskal-Wallis Rank Sum tests for users' View Importance ratings per round. The view importance were reported by users for each decision. In Round 1, we find a significant difference in user view importance rating for Tweet Panel and Entities views. As indicated in Table \ref{fig:summary}, we find the largest difference between the Control group. This result is interesting as it demonstrates that, without intervening on this group with a visual anchor, users value the two supplemental views more than users who may be ``anchored'' to focus only on the two primary and more "visual" views. In Round 1, we find the treatment groups do have a significant effect on the value users rate the Account view, suggesting that such a visual anchor drove users to leverage more of that view for their decisions. However, we do not find a similar effect of the SN view as all three groups (including Control) ranked that view nearly identically throughout the sessions.

\begin{table}[]
\centering

\begin{tabular}{|l|l|c|c|}
\hline
\multicolumn{1}{|c|}{\textbf{Test}}                      & \multicolumn{1}{c|}{\textbf{Outcome}} & \textbf{Round 1}                                              & \textbf{Round 2}                                                     \\ \hline
Chi-Squared                                              & Accuracy                              & \begin{tabular}[c]{@{}c@{}}0.2867\\ (0.8664)\end{tabular}     & \begin{tabular}[c]{@{}c@{}}1.8056\\ (0.4054)\end{tabular}            \\ \hline
\begin{tabular}[c]{@{}l@{}}One-way \\ ANOVA\end{tabular} & Time Spent                            & \begin{tabular}[c]{@{}c@{}}3.0419 *\\ (0.0579)\end{tabular}   & \begin{tabular}[c]{@{}c@{}}0.0879\\ (0.9161)\end{tabular}            \\ \hline
\begin{tabular}[c]{@{}l@{}}One-way\\ ANOVA\end{tabular}  & Confidence                            & \begin{tabular}[c]{@{}c@{}}8.1136 ***\\ (0.0004)\end{tabular} & \begin{tabular}[c]{@{}c@{}}11.8 ***\\ (\textless 0.0001)\end{tabular} \\ \hline
\end{tabular}
\caption{Univariate statistical tests (Chi-Squared and One-way ANOVA) for treatment effects on three participant outcomes (Accuracy, Time Spent, and Confidence) by round. Primary value is statistic value, p-value is in parenthesis. * = 90\% Confidence, ** = 95\% Confidence, *** = 99\% Confidence.}
\label{tab:rq1}
\end{table}

\begin{table}[]
\centering

\begin{tabular}{|c|c|c|}
\hline
\textbf{View Importance} & \textbf{Round 1}                                                        & \textbf{Round 2}                                              \\ \hline
Accounts                 & \begin{tabular}[c]{@{}c@{}}8.3964 **\\ (0.0150)\end{tabular}            & \begin{tabular}[c]{@{}c@{}}3.5761\\ (0.163)\end{tabular}      \\ \hline
Social Network           & \begin{tabular}[c]{@{}c@{}}0.1642\\ (0.9212)\end{tabular}               & \begin{tabular}[c]{@{}c@{}}5.3034 *\\ (0.0705)\end{tabular}   \\ \hline
Tweet Panel              & \begin{tabular}[c]{@{}c@{}}46.779 ***\\ (\textless 0.0001)\end{tabular} & \begin{tabular}[c]{@{}c@{}}3.7133\\ (0.1562)\end{tabular}     \\ \hline
Entities                 & \begin{tabular}[c]{@{}c@{}}22.509 ***\\ (\textless 0.0001)\end{tabular}  & \begin{tabular}[c]{@{}c@{}}13.459 ***\\ (0.0011)\end{tabular} \\ \hline
\end{tabular}
\caption{Univariate statistical tests (Kruskal-Wallis rank sum test) for treatment effects on Likert Scale View Importance Ratings (7 = Extremely Important, 1 = Unimportant). Primary value is statistic value, p-value is in parenthesis. * = 90\% Confidence, ** = 95\% Confidence, *** = 99\% Confidence.}
\label{tab:view-importance}
\end{table}

\begin{table}[t]
  \centering
    \includegraphics[width=1.0\columnwidth]{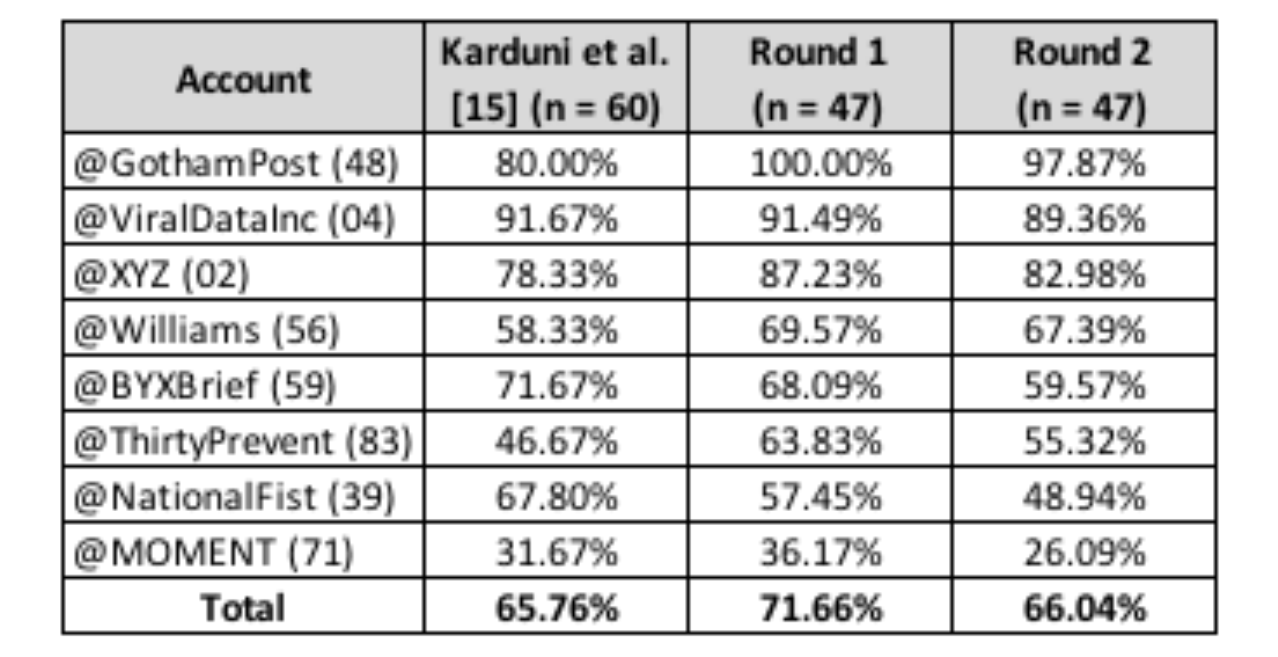}
  \caption{User accuracy by account. Results includes \cite{karduni2018icwsm} which used an earlier version of the Verifi system.}
  \label{tab:accounts}
\end{table}

\textbf{Multivariate analysis.} One weakness of univariate statistical tests is that it ignores relationship among multiple other variables that may also affect accuracy or confidence. To assess such effects, we consider multivariate regression to explain both accuracy and confidence.\footnote{We did not investigate total session time due to the problem of allocating time to each actions for each decision. Therefore, we only investigate accuracy and confidence as dependent variables within a regression framework.} As mentioned earlier, we did not find that the treatments had an effect on accuracy in Round 1. However, similar to our previous study on confirmation bias \cite{karduni2018icwsm}, we find that a more important factor in explaining accuracy is in the difficulty of each account. Table \ref{tab:accounts} provides user accuracy by each account. We find some accounts (e.g., @GothamPost and @ViralDataInc) are very easy for all users and have 90\%+ accuracy. Alternatively, other more difficult accounts -- like @NationalFist and @MOMENT -- had a much lower user accuracy as these accounts had misleading cues or incomplete information (e.g., @NationalFist was not connected on the social network). This implies account-level variation that can be controlled for as a random effect, rather than a fixed effect. As users provided multiple responses, we assume that those responses may be related given they were generated by the same individual who may be better or worse at prediction. Similarly, we also treat each user as a random effect as well. 

\begin{table}[h]
  \centering
    \includegraphics[width=1.0\columnwidth]{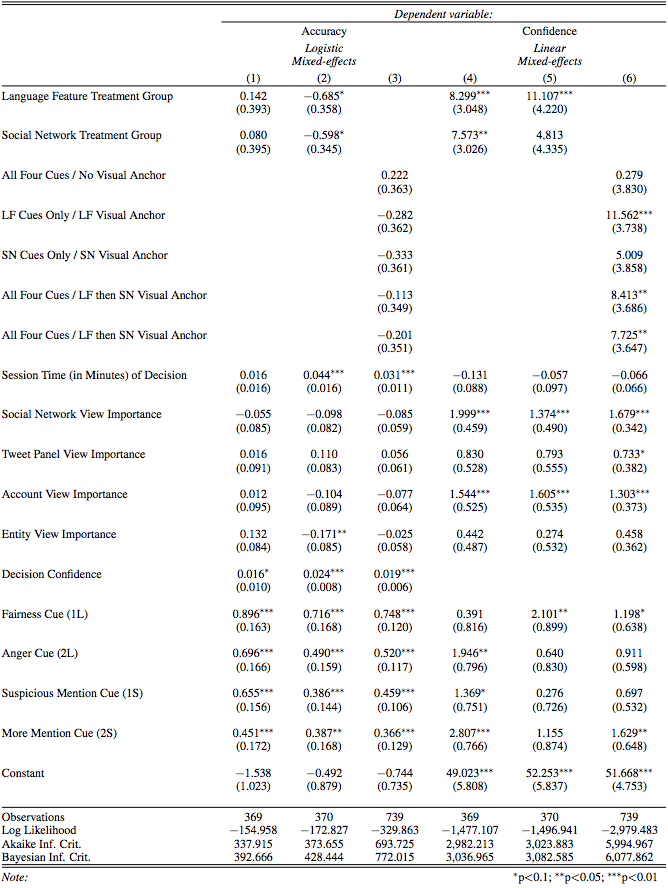}
  \caption{Mixed effects models to explain user accuracy and confidence levels. Reference level for Treatment Group is the Control Group. Reference level for the Cues-Visual Anchor variable is No Cues / No Visual Anchor. Models (1) and (4) are Round 1. Models (2) and (5) are Round 2. Models (3) and (6) are both rounds.}
  \label{tab:regressions}
\end{table}

To consider both account and user-level as random effects, we use a generalized linear mixed effects modeling approach \cite{loy2017model} for each of the two outcome values using the R package {\fontfamily{qcr}\selectfont{lme4}}. For each regression, we use a slight variant depending on the outcome variable format. For accuracy, a binary 1 (correct) or 0 (incorrect) variable, we use a logistic mixed effects model. Alternatively, confidence is a continuous variable between 0 (no confidence) to 1 (perfect confidence) and, hence, we use a linear mixed effects model. 

For each model, we consider ten fixed effects including the four view importance and four strategy cue ratings. One key modification is that for the strategy cues we modified the raw values (1 = yes, 0 = no) dependent on whether the user's cue rating was consistent with the account's actual veracity. For example, cues 1L, 1S, and 2S were phrased so that ``yes'' responses point to real news accounts. Therefore, the modified values are 1 when the cue aligns to the cue's direction, In addition, we include a time of decision variable to attempt to measure potential learning effects (i.e., decisions earlier in the session tend to be less correct or confident than decisions near the end). Table \ref{tab:regressions} provides the regression results for each dependent variable.

\begin{table}[h]
  \centering
    \includegraphics[width=0.9\columnwidth]{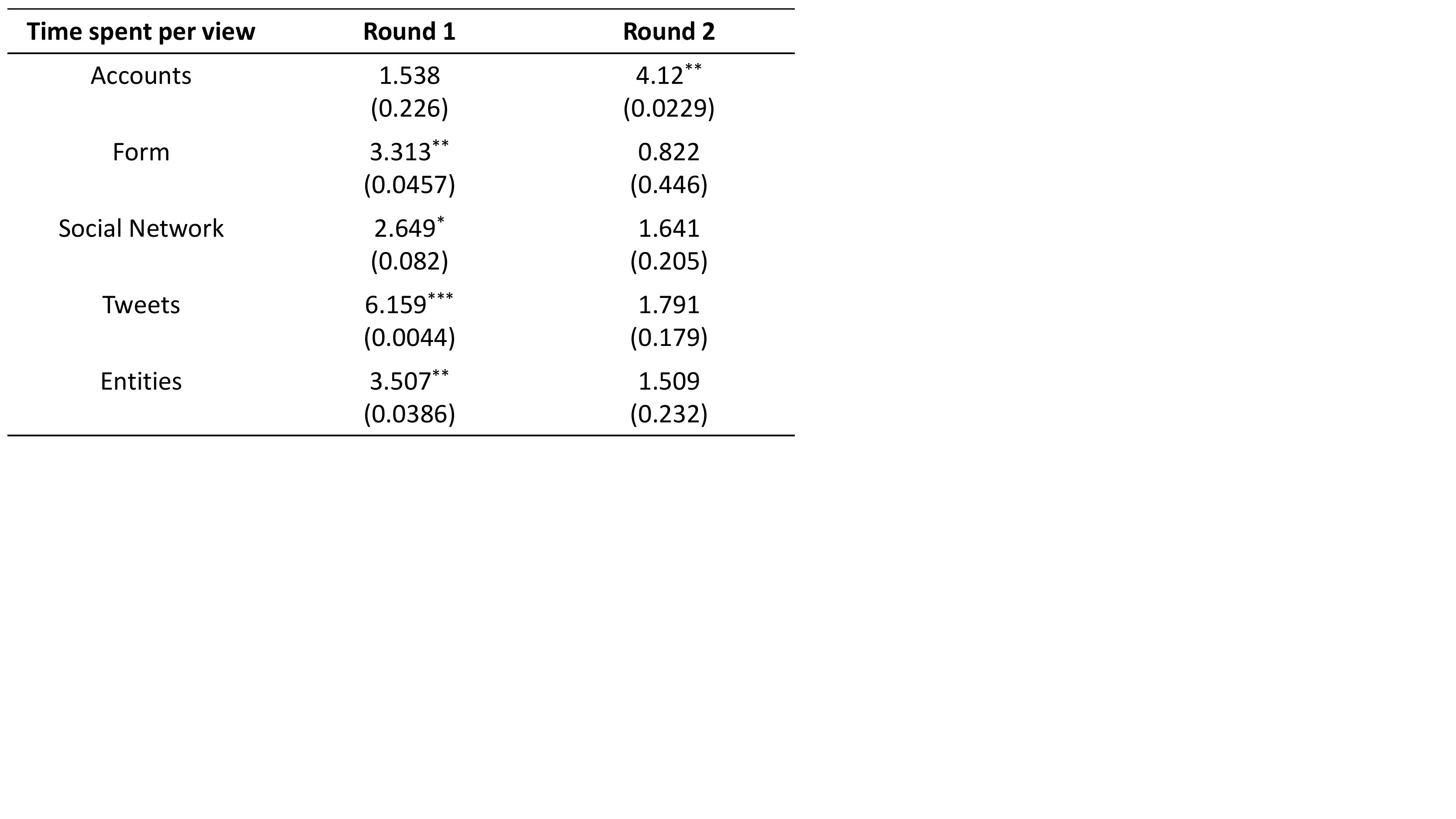}
  \caption{Results from ANOVA statistical test on the time spent per each view between the three groups. * = 90\% Confidence, ** = 95\% Confidence, *** = 99\% Confidence.} 
  \label{tab:time_anovas}
\end{table}

\begin{table}[h]
  \centering
    \includegraphics[width=0.9\columnwidth]{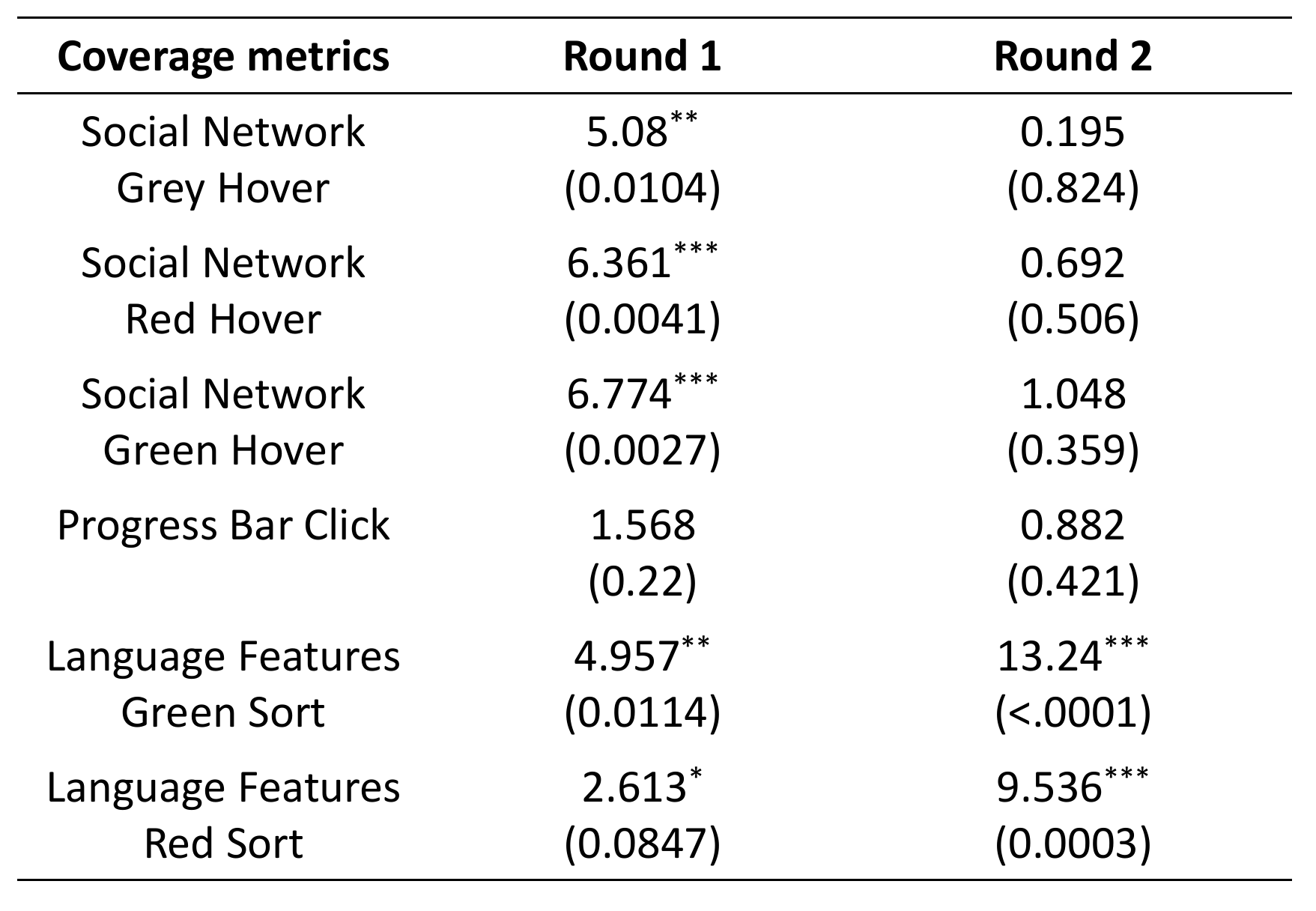}
  \caption{One-way ANOVA tests on the different coverage metrics per treatment group and round. * = 90\% Confidence, ** = 95\% Confidence, *** = 99\% Confidence.}
  \label{tab:coverage_anovas}
\end{table}

In Round 1, the treatment groups had a significant effect on user confidence but not accuracy. Regressions (1) and (4) provide the Round 1 results. After controlling for other variables, we find that users are much more likely to provide higher confidence in the two treatment groups (LF and SN) as compared to the control group. However, the same variables show no significance when predicting users' accuracy. This suggests that either the cues or visual anchor may give users more trust in the system but do not materialize into actual decision-making gains for determining misinformation. We also find that the strategy cues were very important to users' accuracy. All four cues were statistically significant (99\%+ confidence) to explain accuracy. Moreover, coefficients can help rank which cues are more important. For example, the Fairness Cue (1L), when used consistent to the account of interest, had the most significant positive effect on correct responses. Alternatively, the More Mentions Cue (2S) was positively linked when used consistently but to a less extent than the other cues. Last, we find that users' view importance ratings are positively linked to users' confidence levels but not their accuracy. For instance, users with higher Social Network, Tweet Panel, or Account view importance provide higher overall confidence for their decision but such self-assessed ratings do not translate into better decision-making.

\subsubsection{Experiment Results - RQ2}

To answer RQ2 regarding anchoring effect on the analysis process, we consider both univariate statistical tests to identify differences in interaction logs as well as clustering users based on primary and secondary actions and time differences to categorize them based on inferred behaviors and strategies.

\textbf{Time spent per view.} In Round 1, we find certain differences between time spent on each view. To measure time spent per view, we calculated the difference in time between each sequential pair of actions. We then attributed that difference to the later action and, thus, the related view for that action. For example, if a user logged in at time zero, then their next action was a SN node click, we attributed the one second difference to the SN view. We then aggregated total time per view for each user.

To consider statistical differences in time spent, we used one-way ANOVA (with Tukey HSD adjustment) to compare how the treatments may have affected time spent. Table \ref{tab:time_anovas} provides the results of the statistics tests. In Round 1, we find significant differences in the amount of time spent per view in the Form Submit, Tweet Panel, and Entities views. Specifically, we find most differences are between the control group and the two treatment groups rather than differences between the treatments. For example, in the Tweets Panel view, post hoc comparisons between groups indicate significant difference between SN and Control groups ($p=0.0257$) and between the LF and Control groups ($p=0.0047$). Moreover, we find similar, but weaker, significance between time spent in the Entities and Form Submit views. Post Hoc comparisons between the groups indicate significant difference on the Entities view between the LF and Control group ($p=0.0388$).

\textbf{Time to first decision.} In addition to Time spent per view, we also consider the time until each users' first form submit. The goal of this metric is to measure how long the user explored the interface before formally starting his or her decision-process. In Round 1, we found significant differences between time to submit their first form between the treatment groups ($F(2,44) = 4.144 , p = 0.0224$). Specifically, post hoc comparisons between the groups indicate significant difference between the Language Features and Control groups ($p=0.0166$).


\textbf{Coverage Metrics.} In addition to time spent per view, we also created several coverage metrics \cite{wall2017warning} to explore usage of key functionality in the interface. Specifically, we consider six primary actions: progress bar click, LF sort (combined for red/green features), and SN hovers (for grey, green, and red accounts).\footnote{We removed hovers less than one second after a previous action to remove unintentional actions.} These six actions can be categorized as four possible strategies:

\begin{itemize}
    \item \textbf{Language Features:} For this strategy, we measure the time spent on the Accounts view as well as the Language Sort clicks for either the ``green'' (positively correlated with real accounts) or ``red'' (negative correlated with real accounts) features.
    \item \textbf{Social Network:} For this strategy, we measure the time spent on this view as well as the three primary actions related to the social network: hovers on grey, red, and green accounts. To remove unnecessary noise, we removed all hovers committed less than one second to any previous action.
    \item \textbf{Organized:} To measure this, we include Progress Bar clicks to track users who use this functionality to maintain their progress. 
    \item \textbf{Explorer:} i.e., user who takes much longer before moving into decisioning via form submissions. To explore this behavior, we use the time until their first form submit as an additional feature.
\end{itemize}

Table \ref{tab:coverage_anovas} provides the one-way ANOVA tests for each of these metrics. In Round 1, we find significant differences in the use of the social network hover, especially the red-green hovers. These hovers tend to indicate exploring by example -- for example, learning how well connected other known red or green accounts. Using post hoc Tukey tests, we find that there is significance between LF and Control groups ($p=0.0033$) for hovering over the red accounts and between LF and Control groups ($p=0.0018$) for hovering over the green accounts. We also find  strong significant difference in using the LF sort functionality for the green accounts and post hoc comparisons indicate strong significance between the SN and Control groups ($p=0.0109$). We also find weak significance on the LF sort functionality for the red accounts with post hoc comparison showing difference between the SN and Control groups ($p=0.087$).

\begin{figure*}[t]
  \centering
    \includegraphics[width=1.0\textwidth]{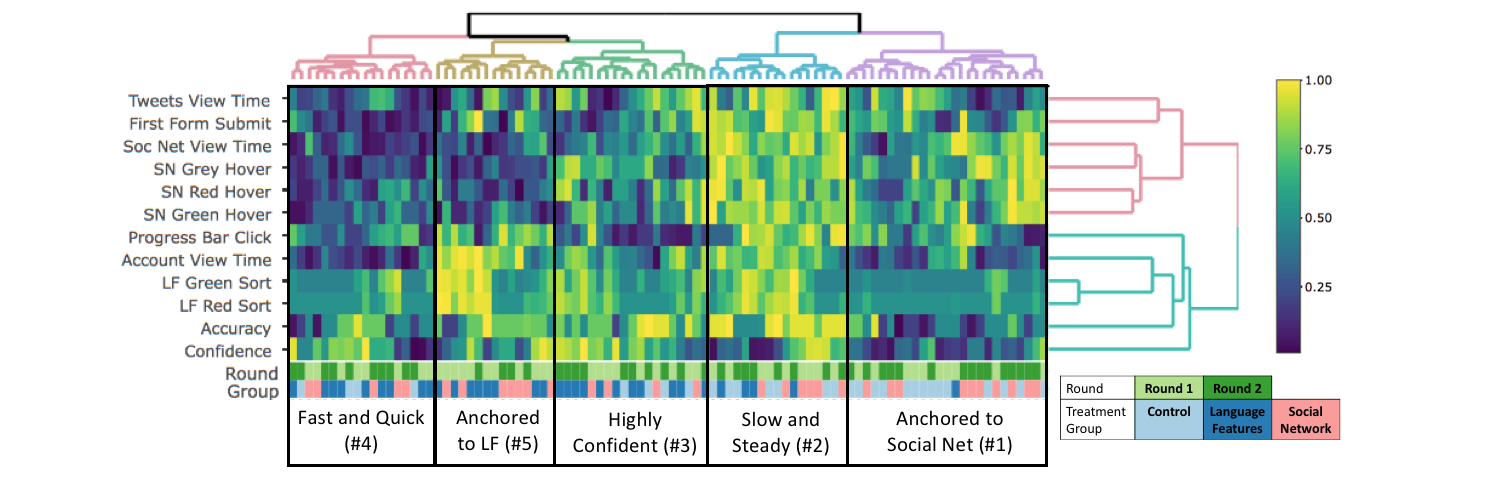}
  \caption{Heatmap clustering of interaction logs (Ward.D2) by columns (users) and rows (metrics). Each column is normalized for its percentile ranks. Users with a high rank of that feature are yellow while users with a low rank usage are dark blue. The bottom two rows indicate user's treatment group and round. Both of these metrics were not used in clustering and provided for comparison.}
  \label{fig:heatmap}
\end{figure*}

\begin{figure}[h]
  \centering
    \includegraphics[width=1.0\columnwidth]{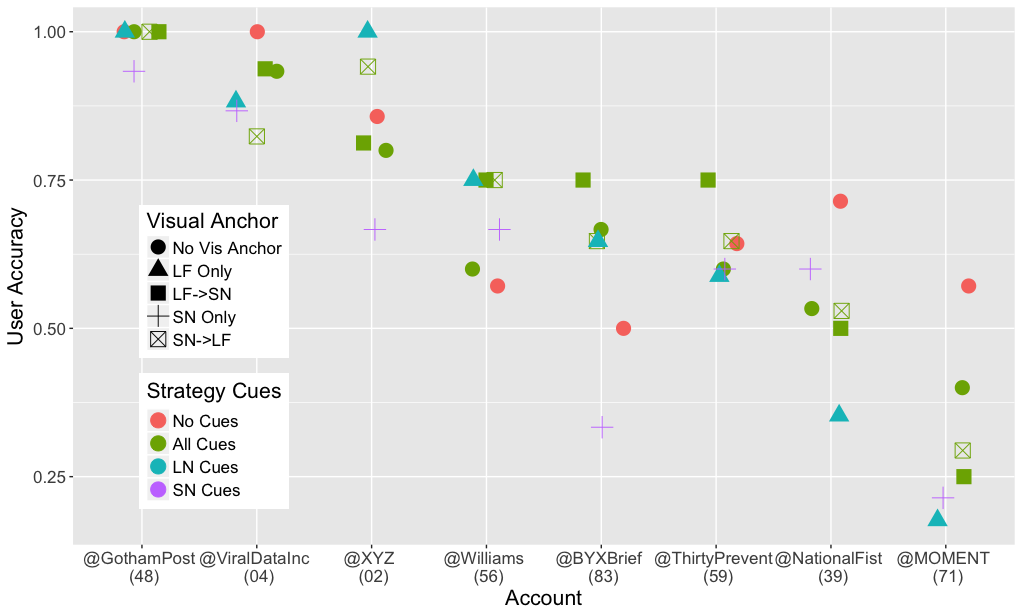}
  \caption{User Accuracy by account (x-axis) and treatment (color and shape). Account order is ranked from left (highest accuracy) to right (lowest).}
  \label{fig:account-perf}
\end{figure}

\subsection{Round 2: One Anchor and Partial Cues}

\subsubsection{Experiment Setup}

Round 2 was motivated to identify the individual effects of each visual anchor and corresponding strategy cues. For example, Round 1 treatments received all four cues as well as two visual anchors -- simply in reverse order. However, from Round 1 it is not clear what is the effect of either cue pairs or visual anchors given the groups received both treatments. To address this problem, we devised Round 2 to build off of Round 1's design but provide partial cues and visual anchors. As in Round 1, all treatment and control groups still received the same general video to introduce all views and functionality. The difference in Round 2 treatment groups is that the SN group only received the SN cues (1S, 2S) and SN scenario video as the visual anchor.  Alternatively, the LF group only received the two LF cues (1L, 2L) and the corresponding LF visual anchor. For the control group, we provided users all four cues (but no visual anchor) to differ from the Round 1 control group in which participants did not receive any cues.

\subsubsection{Experiment Results - RQ1}

\textbf{Univariate analysis.} In Round 2, we find evidence that the treatments had an effect on confidence but not time spent or accuracy (see Table \ref{tab:rq1}). Notably, we find again that the control group had a much lower average confidence than both of the treatments that provided only one set of cues and visual anchors. Nevertheless, we did find that the Round 2 Social Network group had a marked decline in accuracy as shown in Table \ref{tab:rq1}. Regarding View Importance, in Round 2 we find a significant difference in user view importance for Entities view and a slight difference for the Social Network view.

\textbf{Multivariate analysis.} To analyze user accuracy and confidence, we again employ  mixed effects models on Round 2 (regressions (2) and (5)) and then combine both rounds (Table \ref{tab:regressions}). For accuracy, we find a slight effect of SN treatment group in Round 2, as that group's performance was the lowest out of any round-group treatment. Moreover, in Round 2, we observe a learning effect as the time of the users' decision is positively related with higher accuracy. Like Round 1, we also find that consistent strategy cue use are strongly correlated with accuracy. This observation indicates that, holding all variables constant, users who performed much better when they correctly employed the strategy cues. Once again, the Fairness Cue (1L) is the most important as it has the largest coefficient value. Last, like Round 1, we find higher confidence levels tend to be positively related to more accurate decisions but with a higher level of statistical significance.

Alternatively, in the explanation of users' confidence in Round 2 (i.e., regression (5)), we find that the Language Features treatment group is positively associated with nearly an 11 point higher confidence level than the control group for Round 2. In addition, we find that users' view importance ratings for Social Network, Tweet Panel, and Account views are positively related to confidence like Round 1. 

Last, to isolate specific treatment effects, we combined both rounds to create regressions (3) and (6). In these models, instead of using the treatment groups as a covariate, we combined them to create a six-level treatment variable with the reference is the Round 1 control group (no cues / no visual anchor). In these models, we find that neither providing cues nor the visual anchors have a statistically significant fixed effect on user accuracy. Like previous rounds, strategy cues (when used), time and user confidence affect user accuracy. This suggests that while the strategy cues were helpful, some individuals choose to ignore (or perhaps did not fully trust or understand) the strategy cues. Second, we find visual anchors, especially the language features, have a positive effect on user confidence. For example, as compared to having no visual anchor (or cue), users provided both visual anchors on average had nearly an eight point higher confidence score. Interestingly, the social network cue alone does not provide a similar gain. One possible explanation could be users felt more comfortable with the social network views originally and hence, additional reinforcement of this view and strategy did not add further confidence.

\textbf{Account-level analysis.} While we did not find that the treatments had, on average, an effect on accuracy, we find that the treatment groups have some variation in accuracy when controlling for the account. Figure \ref{fig:account-perf} provides the accuracy for each treatment as encoded by color (strategy cues provided) and shape (visual anchor). The x-axis provides decisions for each each account by treatment and the y-axis provides that treatment's accuracy. Slight x-axis jittering was provided to separate points. Consider @MOMENT (Account 71) which was the most difficult account. We observe that groups with no visual anchor performed better on this account. The issue with Account 71 was that its cues were conflicting as it was only connected to a red (suspicious) account whereas its language features were not entirely consistent with 1L and 2L. Because of this problem, not only did the cues tend to hurt performance, but even visual anchors seem to drive sub-optimal performance for this account. 

\subsubsection{Experiment Results - RQ2}

Similar to Round 1, we find differences in Time Spent (Table \ref{tab:time_anovas}) per view and coverage metrics (Table \ref{tab:coverage_anovas}) in Round 2 depending on the treatments provided. We find there is strong significant differences on Time Spent between the SN and LF group ($p=0.0372$) on the Accounts view. On the coverage metrics, we find significant difference on the use of LF sort between the red (suspicious) and green (real) accounts. Post hoc comparisons show a strong significance between the LF and Control groups ($p=0.0008$) and between SN and LF groups ($p=0.00006$) on the usage LF green sort. We also significant difference between the LF and Control groups ($p=0.0033$) and between the SN and LF groups ($p=0.0007$) while using the LF red sort.  

\textbf{Clustering users based on their interactions.} To identify user behaviors with the coverage and time spent metrics, we used Ward's D2 Agglomerative Hierarchical Clustering \cite{murtagh2014ward} to cluster users and features. To improve our results, we combine both Rounds 1 and 2 instead of running clustering on each individual round. Figure \ref{fig:heatmap} provides a cluster heatmap visualization from the R package {\fontfamily{qcr}\selectfont{heatmaply}} \cite{galili2017heatmap}. In this figure, each metric is normalized as the percentile (rank) across the metric. For example, yellow indicates a user who ranks high on that metric compared to all other users (regardless of group or round). Dark blue represents users with a low rank of that metric.

\begin{figure}[h!]
  \centering
    \includegraphics[width=1.0\columnwidth,keepaspectratio]{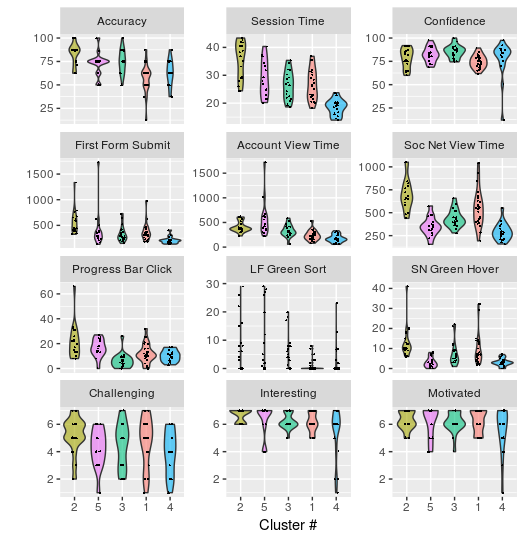}
  \caption{Violin plots by cluster groups using R {\fontfamily{qcr}\selectfont{ggplot2}} \cite{wickham2016ggplot}. Points are on a user-level. Each row of charts is by metric category (e.g., primary, time (in sec.), coverage, post-questionnaire). Slight x-axis jittering was added for point visibility. \#1 = Anchored to SN, \#2 = Slow and Steady,\#3 = Highly Confident, \#4 = Fast and Quick, \#5 = Anchored to LF.}
  \label{fig:cluster-stats}
\end{figure}

\begin{figure*}[h]
  \centering
    \includegraphics[width=1.0\textwidth]{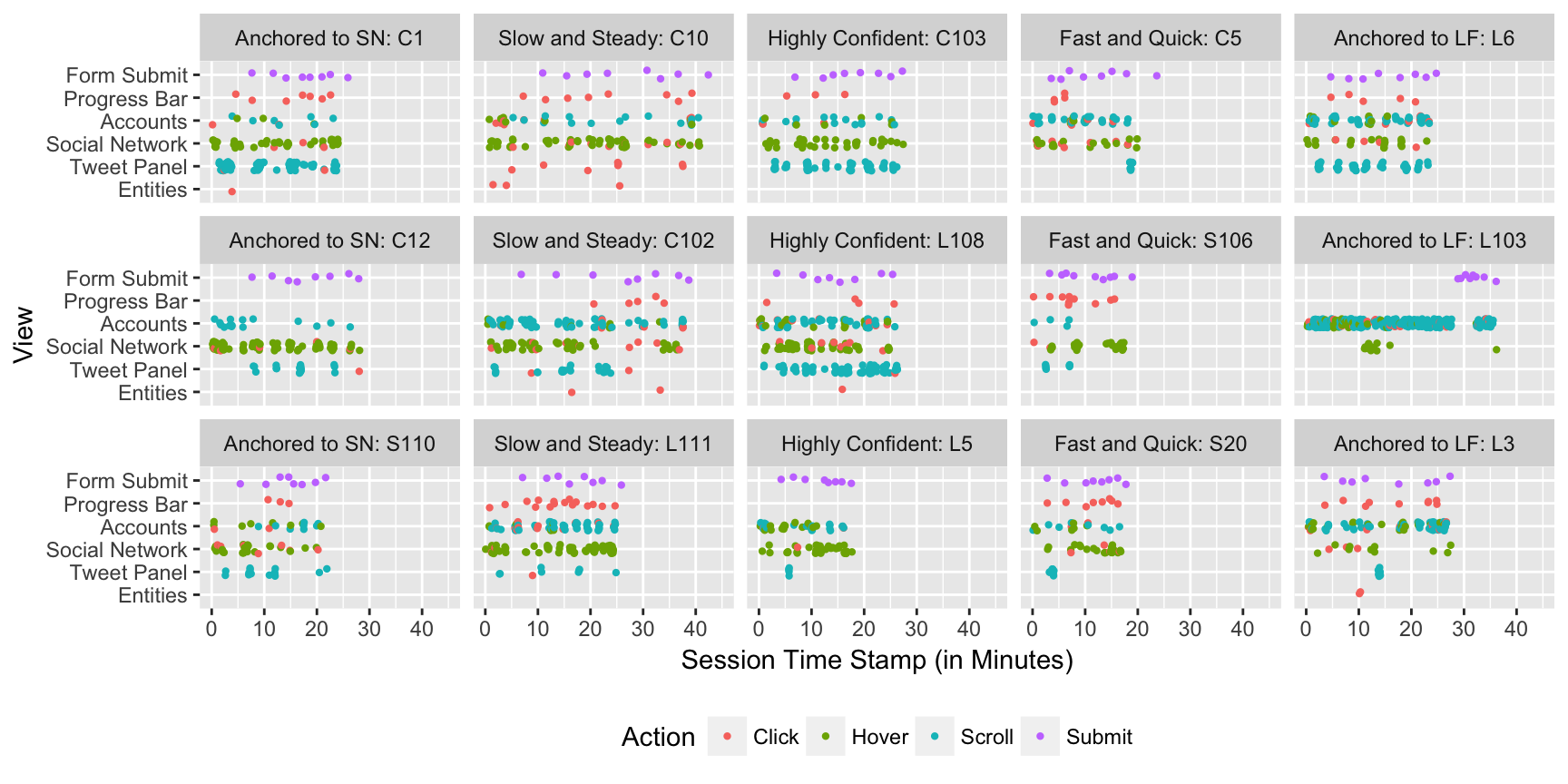}
  \caption{Experiment interaction logs of Verifi using R {\fontfamily{qcr}\selectfont{ggplot2}} \cite{wickham2016ggplot}. Each plot is a user's interaction log. Each dot is a user action: click (red), hover (green), scroll (blue), and submit (purple). The x-axis is the time of the action. The y-axis is the respective view associated with that action. The order corresponds to critical functionality (e.g., Form Submit) to primary view (e.g., Accounts vs. Social Network) to secondary views (e.g., Tweet Panel or Entities). Slight y-axis jittering has been applied on the view level to avoid point overlap. Chart columns indicate user-level strategies based on user-level dendrogram clustering. Chart row order represents, in descending order, highly accurate users (7+ out of 8, top row), average users (5-6 out of 8, middle row), and inaccurate users (4 or less of 8, bottom row).}
  \label{fig:userlogs}
\end{figure*}

To determine the optimal number of clusters for the rows (features) and columns (users), we used the maximal average silhouette width method on the cophenetic distance of the dendrogram \cite{galili2015dendextend}. The algorithm detected five clusters on the user-level, as identified by the five colors in the horizontal dendrogram. We then annotated the five clusters based on common attributes shared by users within a cluster. 

We find the clusters can infer user strategies. For example, the `Slow and Steady' cluster is mostly yellow, indicating a high rank across all metrics. These users explored the entire interface's functionality for an extended period of time. On the other hand, the `Fast and Quick' group is mostly dark blue as they ranked low in coverage metrics and time spent. The bottom two rows of the dendrogram provide the treatment group and round information for each user. One hypothesis is that users anchored on different view would adopt different analysis strategies. If this were true, we would expect that users would cluster based on such treatments. In part, we find some evidence. Take `Anchored to Social Network' group as an example. Only one user who was treated with a LF visual anchor (dark blue) within this cluster. As we would expect, many are SN groups (light red) that received the SN visual anchors. However, what's peculiar is the number of Control users (light blue), particularly those from Round 1 (light green). These users were not even given the social network cues! These users seem to naturally be drawn to this view more than other views.

Descriptive statistics and distribution plots (Figure Figure \ref{fig:cluster-stats}) can also provide more context on each cluster. We find that the `Slow and Steady' cluster users averaged much longer session times (M = 35.97 minutes). Further, these users were late starters who explored early. They averaged nearly 10 minutes before each users' first decision submission. For context, other groups typically made their first decision between 3 and 7 minutes. We also find that these users actively used the Progress Bar (M = 21.5 times), indicating more organization, while also using both primary views (Social Network and Accounts) frequently. Interestingly, this cluster has, on average, the highest accuracy of 82.8\%. Alternatively, we identified two clusters as users who focus more on either the SN (\#1) or LF (\#2). For example, cluster \#1 spent 2.3x more time on the Social Network view than the Account view (i.e., LF) whereas the opposite holds for cluster \#2.  

Last, we validated the clusters using response and post-questionnaire data that was not included in the clustering process. For instance, we find that the clusters provide a range of different ratings for the language features and social network functionality in the post-questionnaire. Users in the `Anchored to Social Network' (\#1), `Highly Confident' (\#3), and `Fast and Quick' (\#4) generally preferred the social network over the language features. However, the `Anchored to Language Features' cluster (\#5) was the only cluster to prefer, on average, the LF over SN. Alternatively, we can find distinct differences in user motivation, interest, and challenge between clusters like `Slow and Steady' (\#2) and `Fast and Quick' (\#4). The `Slow and Steady' cluster tended to be the most motivated, interested, and challenged out of all of the clusters. This makes sense given their longer session times and heavy usage. While on the other hand, the `Fast and Quick' cluster was the least motivated and interested. Likely this lack of interest led to their shorter session times and may factor in their lower accuracy.


We also used visual analytics to explore the user-level interaction logs by clusters. Figure \ref{fig:userlogs} provides a scatter plots of fifteen example user sessions.  In each plot, a dot represents an action for each of the six views. Slight y-axis jittering is applied to spread out overlapping actions. The x-axis represents the session time (in minutes) of each individual action. Each chart column represents three example user sessions from each cluster. Chart row order represents, in descending order, highly accurate users (7+ out of 8, top row), average users (5-6 out of 8, middle row), and inaccurate users (4 or less of 8, bottom row). Users C10, C103, and L6 had 100 percent accurate while S110 had the worst accuracy (1 out of 8). Outlier behaviors can also be identified from this plot. For instance, L103 followed the LF cues by almost exclusively using the Accounts view. Moreover, this user waited until the end of the session to make all decisions. 

We were able to identify general patterns from these plots too. For example, the left-most column provides three users who are clustered to the `Anchored to Social Network' group. These users tend to have many more actions in the Social Network view as compared to the Accounts, Tweet Panel, or Entities view. They seldomly use the Progress Bar (e.g., S104 and C1 use it somewhat while S108 never used the Progress Bar). Alternatively, we find examples in the `Slow and Steady' group to have much longer user sessions, lasting well over thirty minutes (some even near forty minutes or more). These users tend to use a combination of all views like the Accounts, Social Network, and even the Tweet Panel views. 

\textbf{Post-Questionnaire Feedback.} Additional insight in understanding user strategies can be gleaned from the qualitative feedback provided by users in the post-questionnaire. For instance, some participants identified a lack of trust in the language features because of a lack of clarity of their composition: ``I did not like making a decision based on you saying whether the language measures were good or bad, I wanted to understand the language measures better.'' Others commented on the need for additional interface features, like a help menu, to aid in this intensive cognitive process: ``it would be beneficial to have a `help' section ON the platform to look at when needing the reminder of things the video mentioned.'' Other users commented on the usability of views in general, like the entities and Tweet Panel view. For example, one user commented ``I didn't really understand the need of entities to determine fake articles.'' While another user admitted that ``I did not use the tweets or entity features of the interface.'' Both comments explain users' limited use of that view but was expected given the limited training to functionality for these views.

\section{Implications for VA Evaluation Practices}
We argue that our findings are informative for guidance on training and tutorial during visualization evaluation with human subjects. Our findings show that visual anchors and strategy cues can significantly impact users' confidence and time spent investigating in each view when performing tasks. In addition, evidence from our study suggest that being anchored to a particular view (SN) can lead to significant worse accuracy (Round 2). Anchoring to a subset views would lead to the over-reliance on (often incomplete) information presented in those views, thus preventing users from getting a comprehensive picture. 

Such anchoring could occur due to the way we train participant how to user the visual interface before asking them to carry out the tasks. First, providing a general training video is a good idea, however, careful considerations are needed when devising a script or training video. The experimenter may want to make sure that all important features/views get equal coverage in the script/video.
Second, providing a secondary video/script walking participants through solving the task with an example dataset is a great way to help participants get started. However, experimenters may unknowingly anchor some participants on an implied strategy implemented in the video/script. 

Since our experiments show that visual anchors can indeed impact multiple performance metrics (confidence, accuracy, time to decision), we would like to raise awareness of participants possibly being unintentionally anchored and suggest careful consideration on how to train users to use a visual interface.

\section{Limitations and Future Work}


In this section, we outline study limitations along with identifying areas of future work for analyzing cognitive biases in visual analytics.

\textbf{Limitations.} One limitation of our study was limited testing on the design of the interface. While the training process differed between groups, all users received the same interface. However, design layouts could have interaction effects with treatments. One approach could provide revised interface layouts to identify the marginal value of design or even each specific view in the decision-making process under cognitive bias treatments. For example, testing whether the strategy cues with only the Tweet Panel view (i.e., mimics everyday social media usage) can measure a baseline accuracy. With such a baseline, a more precise estimate of the effect of the visualizations can be inferred.

A second limitation is the choice of accounts in the decision-making task. If we were to have selected more difficult accounts (like @MOMENT) than easy accounts (e.g., @GothamPost), we may find cognitive biases have a larger effect. Moreover, different accounts may also lend to other strategy cues that could affect the treatments.

\textbf{Future Work.} There are several promising paths of future work for understanding cognitive biases through visual analytics. First, there are many opportunities to expand Verifi to include additional tools in identifying misinformation including images, semantic text analysis (e.g., word embeddings), and account-level clusters. A newer version of Verifi \cite{karduni2018vast} addresses many of these issues and provides a longer dataset of accounts with a broader range of accounts. With different stimuli, future experiments could explore the effect of visual anchors on image exploration (e.g., can exposures to extreme emotions affect users' performance when provided images as well?). Further, future system iterations could include streaming components that test decision-making under dynamic data. 

Second, future systems could incorporate a ``suspicious'' supervised model (e.g., \cite{volkova2017separating}) as a credibility score for decision-makers. This would enable interpretation of higher dimensional features into a single vector. In doing so, users could be ranked on overall (or dimension level). A credibility score would lend itself to combine more cues into a transparent, easy to understand heuristics as cues (e.g., any accounts over score x are suspected of misinformation). 

Last, more research is needed on how individual differences affect decision-making in visual analytics. Our results, while promising, also indicate that some users are not affected by the strategy cues or visual anchors (e.g., some anchored to social network were from a different treatment). Said differently, some users' decision-making seem to be based on their individual traits (e.g., experience, familiarity \cite{pennycook2018crowdsourcing}, cognitive ability \cite{bergman2010anchoring}) rather than treatments. Future work could incorporate more sophisticated experiment designs by attempting to identify heterogeneous treatment effects \cite{salganik2017bit, athey2017econometrics}. 

\section{Conclusion}

In this paper, we presented an experiment on the role of anchoring bias in users' decision-making, interaction paths, and confidence in identifying misinformation on Twitter in visual analytic systems. We find that providing visual anchors and strategy cues can greatly affect users' confidence but have mixed effects on users' speed and decision accuracy. Secondary factors like view importance can also play a role in users' confidence while strategy cues can drastically improve decision-making if used correctly (and not ignored). Last, exploration of user interaction logs can provide hints to users' strategies and the effects such treatments can have for certain individuals. While we find that some users are susceptible to such anchoring biases, others can ignore such treatments -- perhaps due to uncertainty or a lack of trust -- leading individual attributes like motivation or interest can explain more of the users' knowledge seeking behaviors.


\bibliographystyle{abbrv}


\end{document}